\documentclass[12pt]{report}
\usepackage{epsfig}
\usepackage{epstopdf}
\usepackage{graphics}
\usepackage{graphicx}
\usepackage{amssymb}
\usepackage{makeidx}
\usepackage{amsfonts}
\usepackage{amstext}
\usepackage{amsmath}
\usepackage{amsbsy}

\begin{document}
\setcounter{page}{1}        

\title{\sf{
Exclusive Topologies reconstruction\\
 in LAr-TPC experiments:\\
 a Novel Approach for precise \\
 Neutrino-Nucleus Cross-Sections Measurements\\
 \vspace{1.0cm}
{\it \sf\large White Paper}\\
}
\vspace{0.5cm}
{\rm \small {O. Palamara}$^{{\mbox{a}},{\mbox{b}}}$,
{K. Partyka}$^{\mbox{a}}$,
{F. Cavanna}$^{{\mbox{a}},{\mbox{c}}}$\\
for the {\sf ArgoNeuT} Collaboration\\
\vspace{2.5cm}
{\it $^{\mbox{a}}$ Yale University, New Haven, CT (USA)}\\
{\it $^{\mbox{b}}$ INFN - Laboratori Nazionali del Gran Sasso, Assergi, Italy}\\
{\it $^{\mbox{b}}$ Universit\`a dell'Aquila and INFN, L'Aquila, Italy}\\
}
}
\maketitle
\begin{abstract}
{\small The analysis and interpretation of the present and future neutrino oscillation experiments
strongly rely on the quantitative understanding of neutrino and antineutrino interactions with
nuclei in the "few GeV" energy range.
With the advent of the Liquid Argon Time Projection Chamber (LArTPC) experimental
techniques for neutrino detection, a novel approach and methodology in neutrino data analysis 
is now being developed. Purpose of this White Paper is to briefly illustrate this method, 
as based on real events categorization into experimental exclusive topologies, 
rather than on MC driven classification of the event rates in terms of interaction channel.
Data collected by the ArgoNeuT experiment in the "few-GeV" energy range are extremely helpful to directly 
probe features of neutrino interactions and associated nuclear effects in the range of interest for future 
long-baseline neutrino oscillation experiments.}
\end{abstract}


{\sf  The analysis and interpretation of the present and future neutrino oscillation 
experiments strongly rely on the quantitative
understanding of neutrino and antineutrino interactions with nuclei in the "few GeV" energy range.  \\
In this range the CC quasi elastic (QE) scattering is the most important interaction channel. 
This reaction is commonly identified with the emission of a charged lepton and a single proton at the vertex.  
However, with current Cherenkov detectors (C target for liquid scintillator or O target for water) 
emitted protons are below detection threshold and
different other mechanisms or reactions may lead to final states that  cannot be
distinguished from those originating by genuine QE processes 
(and it has been suggested that the excess of CCQE cross section observed by MiniBooNE 
may be traced back to the occurrence of some unconventional processes among these). 
Therefore, the observation of a single lepton signature in final state is often referred to as CC {\it ``QE-like"}, and a lot of experimental effort has  been devoted to measurements of neutrino- and
antineutrino-nucleus QE-like cross-sections in a broad kinematical domain.\\
From the theoretical viewpoint, over the last decade [NuInt Workshop series] the research focused on
the systematic study of the nuclear effects and their impact on the determination of neutrino cross sections in the same "few-GeV" energy region. 

Nuclear effects, including short- and long-range nucleon correlations inside the nucleus (usually, but sometime too generically, indicated by SRC and MEC processes) and intra-nuclear re-scattering processes in the nuclear target (or FSI, final state interactions) play a key role in neutrino-nucleus interactions.
Due to one or to the interference of some of these mechanisms, even the simpler QE interaction can often be accompanied by the ejection of additional nucleons (protons and/or neutrons), 
emission of $\gamma$'s from nuclear de-excitation and sometimes by soft pions
in the final state. Again, these products were so far neglected
because not detectable and the identification of the neutrino interaction channel remains largely ill defined. All  cross-section measurements of specific processes therefore largely rely on MonteCarlo simulation.

With the advent of the Liquid Argon Time Projection Chamber (LArTPC) experimental techniques for neutrino detection, with its full 3D imaging combined with efficient particle identification capability and precise calorimetric energy reconstruction, new perspectives are now open for a detailed reconstruction of the complex event topologies emerging in final state from neutrino-nucleus interactions.
LAr-TPC detectors provide bubble-chamber-like quality images and robust background rejection power,
allowing for {\it exclusive topologies} recognition and direct exploration of nuclear effects  with extraordinary sensitivity. \\
A novel approach and methodology in neutrino data analysis with LArTPC is now being developed, based on the first event sample collected by the ArgoNeuT experiment (Argon Neutrino Test) [1] in the "few-GeV" energy range, of interest for future long-baseline neutrino oscillation experiments. \\
More specifically, {\bf instead of MC based classification of the event rates in terms of interaction channel ({\it QE, RES, DIS, etc}), 
neutrino events in LAr can be directly categorized in terms of final state topology based on particle 
multiplicity}: \\
{\it ``0-pion"} (i.e. $\mu$+Np, where N=0,1,2...), {\it ``1-pion"} (i.e. $\mu$+Np+1$\pi$,  N=0,1,2...) events, {\it etc}..\\

With the imaging LAr-TPC detector these exclusive topologies can be fully reconstructed, 
measurements of proton multiplicity at the neutrino interaction vertex and reconstruction of
 proton kinematics in events with different proton multiplicity can be performed down to very low proton energy threshold.
 In exclusive topology cross-section determination, the use of MC information is limited to the estimate of the event selection efficiency and acceptance (where reliability of detector  simulations is very well established).
 Ratios among rates of different exclusive topologies provide indications of the size of nuclear effects, and possibly constrain their nature.
 Ultimately, this approach will also  allow for most precise reconstruction of the incoming neutrino energy from the lepton {\it AND} proton (and pion) reconstructed kinematics.\\
The ArgoNeuT detector [1], 170 L active volume LArTPC, collected several thousands of CC muon neutrino events during an extended run in 2009-10 with the NuMI LE beam at Fermilab
in both $\nu$-mode ($\sim2$ weeks, 8.5x10$^{18}$ POT, $\langle E_\nu\rangle\simeq3.5$ GeV) and anti-$\nu$-mode (about five months, 1.25x10$^{20}$ POT, 40\% $\bar\nu$ with $\langle E_{\bar\nu}\rangle \simeq3.5$ GeV and 60\% $\nu$ with $\langle E_\nu\rangle\simeq7$ GeV). 
A first topological analysis, for the "0-pion" muon neutrino events [separately for the neutrino ($\mu^-$) and anti-neutrino ($\mu^+$) sample] is currently developed with a proton threshold of 21 MeV Kinetic energy. Preliminary results
were presented at recent conferences (e.g. NUINT 2012 Workshop [2], [3]). A series of ArgoNeuT events as examples of different topologies currently under study are shown in Fig.\ref{events}.

Accurate and detailed MonteCarlo neutrino generators are needed for comparison with LAr data.
Inclusion of a realistic treatment of the nuclear effects in MC codes represents the most difficult challenge in current MC development.
The results of a new generation of theoretical studies are shading new light on the details of neutrino-nucleus
interactions, but also raising a number of outstanding still unresolved issues.
Data from LAr are extremely helpful to give answer to many of the open questions 
and to provide important hints 
(for example from the reconstruction also of the neutrons ejected in the final state)
to tune MC generators and discriminate among theoretical models. 

To this end, progressing with the development of more and more accurate reconstruction tools for LArTPC data analysis is now deemed necessary.  
Due to the high imaging quality of LArTPC, visual scanning represents a powerful
method to directly probe features of neutrino interactions, otherwise
 inaccessible with other technologies and existing experiments.
Visual analysis is propedeutical at this stage to first guide the definition of the best suited methods
for the topological analysis of neutrino events
and will then tremendously help the development of sophisticated automated reconstruction 
algorithms necessary in the future when larger mass and high statistics LAr-TPC detectors  (MicroBooNE and future LAr detectors like LBNE) will come on-line.\\

\begin{figure}[h]
\begin{center}
\includegraphics[width=3.1in]{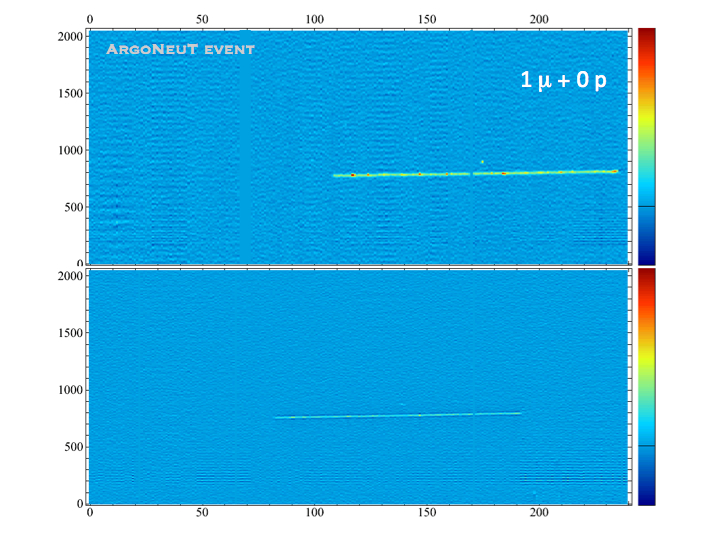}
\includegraphics[width=3.1in]{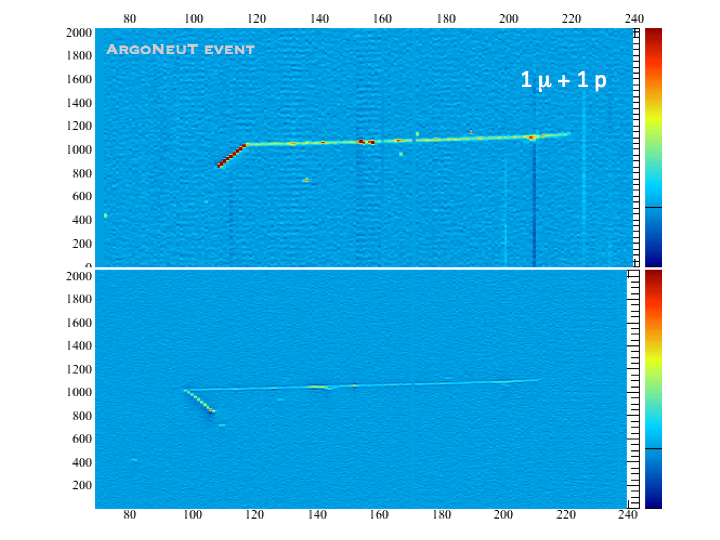}
\includegraphics[width=3.1in]{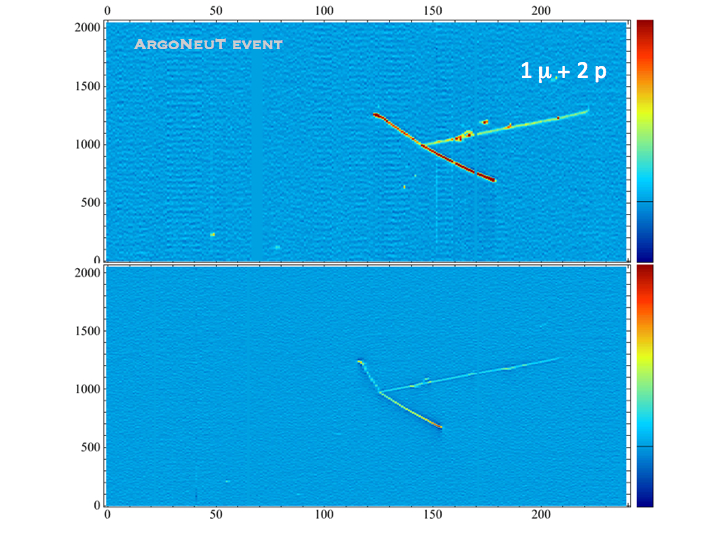}
\includegraphics[width=3.1in]{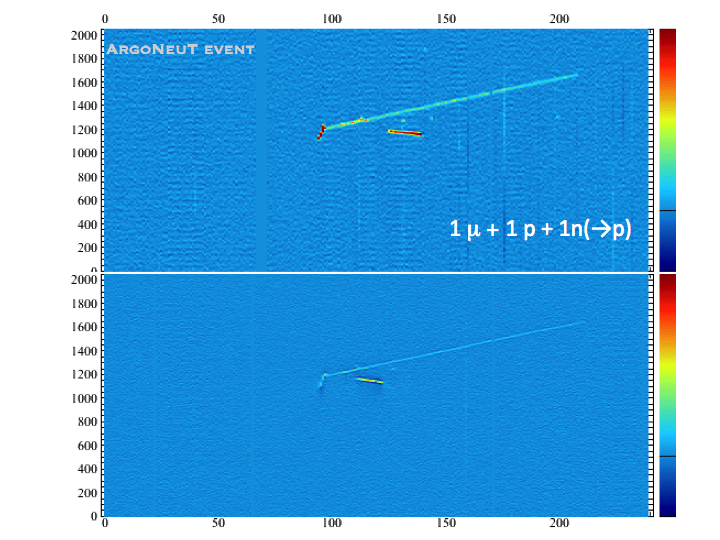}
\includegraphics[width=3.1in]{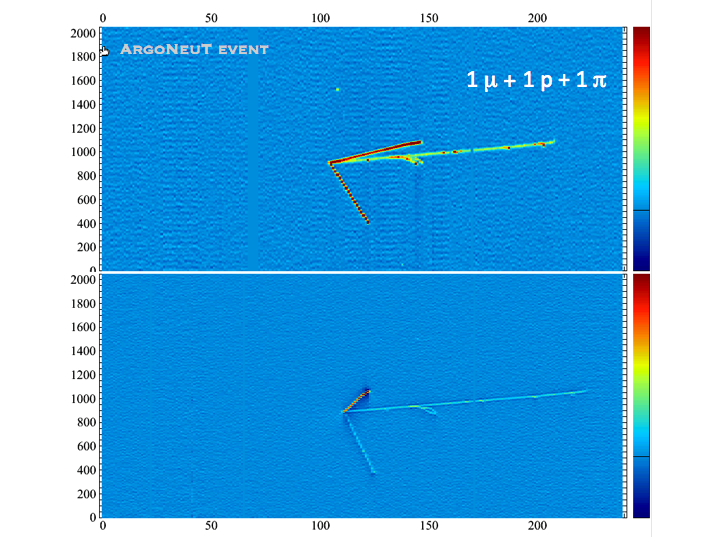}
\includegraphics[width=3.1in]{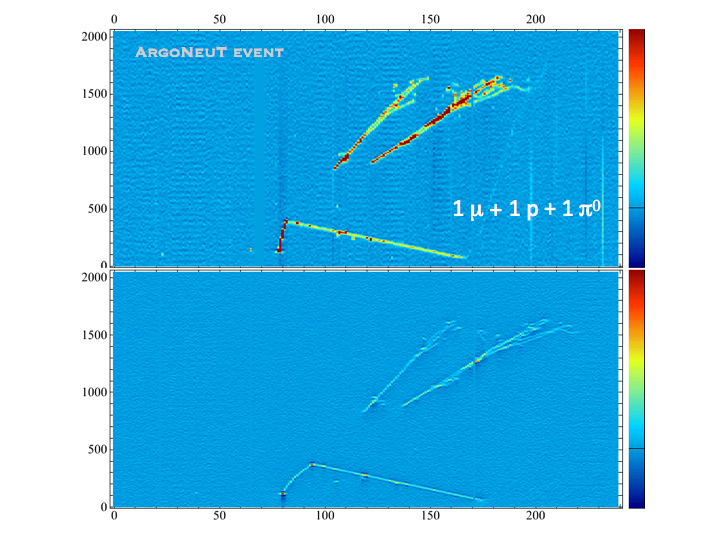}
\caption{\small\sf Selection of ArgoNeuT events examples of different topologies currently under study.
[Top-Left] Single $\mu^-$ event (no proton detected at the vertex),
[Top-Right] Muon with one proton, 
[Center-Left] Muon with two protons (back-to-back),
[Center-Right] Muon with a proton at the vertex and a second proton detached from vertex,  
produced by a neutron from the vertex through charge exchange process ($\mu+1p+1n$ topology),
[Bottom-Right] Muon with one proton and one charged pion (downward going track), 
[Bottom-Left] Muon with one proton and one neutral pion (with $\pi^0\rightarrow 2\gamma$ yielding two well separated el.m. showers).} 
\label{events}
\end{center}
\end{figure} 

}

{\small\noindent [1] ArgoNeuT Coll., {\sf ``The ArgoNeuT detector in the NuMI low-energy beam
line at Fermilab"}, \\
\indent 2012 JINST 7 P10019.\\}
{\small [2] K. Partyka (for the ArgoNeuT Coll.), {\sf  "Exclusive 0-pion topologies in ArgoNeuT".}\\
\indent NuInt12 Conf. Proceedings.\\}
{\small [3] O. Palamara (for the ArgoNeuT Coll.), {\sf "Hints for nuclear effects from ArgoNeuT data".}
\indent NuInt12 Conf. Proceedings.\\}

\end{document}